\documentclass[aps,prd,twocolumn,superscriptaddress,showpacs,amsmath,amssymb]{revtex4}
\usepackage{epsf}
\usepackage{latexsym}
\usepackage{amsmath}
\usepackage{amsfonts}
\usepackage{amssymb}
\usepackage{graphicx}
\newcommand{\be}{\begin{eqnarray}}
\newcommand{\ee}{\end{eqnarray}}
\newcommand{\ud}{\mathrm{d}}

\newcommand{\lp}{\ell_{\rm p}}
\newcommand{\mpl}{M_{\rm p}}
\newcommand{\mew}{M_{\rm ew}}
\newcommand{\meff}{M_{\rm eff}}
\newcommand{\mc}{M_{\rm c}}
\newcommand{\mmax}{M_{\rm max}}
\newcommand{\rem}{R_{\rm EM}}
\newcommand{\reff}{R_{\rm eff}}
\newcommand{\rh}{R_{\rm H}}

\begin{document}
\title{ Possibility of Catastrophic Black Hole Growth in the Warped
Brane-World Scenario at the LHC}
\author{Roberto~Casadio}
\email{casadio@bo.infn.it }
\affiliation{Dipartimento di Fisica, Universit\`a di
Bologna and I.N.F.N., Sezione di Bologna,
via Irnerio 46, 40126 Bologna, Italy}
\author{Sergio~Fabi}
\email{fabi001@bama.ua.edu}
\affiliation{Department of Physics and Astronomy, The University
of Alabama, Box 870324, Tuscaloosa, AL 35487-0324, USA}
\author{Benjamin~Harms}
\email{bharms@bama.ua.edu}
\affiliation{Department of Physics and Astronomy, The University
of Alabama, Box 870324, Tuscaloosa, AL 35487-0324, USA}
\begin{abstract}
In this paper we present the results of our analysis of the growth and decay of
black holes possibly produced at the Large Hadron Collider, based on our previous
study of black holes in the context of the warped brane-world scenario.
The black hole mass accretion and decay is obtained as a function of time, and the
maximum black hole mass is obtained as a function of a critical mass parameter.
The latter occurs in our expression for the luminosity and is related to the size
of extra-dimensional corrections to Newton's law.
Based on this analysis, we argue against the possibility of catastrophic black hole
growth at the LHC.
\end{abstract}
\pacs{04.70.Dy, 04.50.+h, 14.80.-j}
\maketitle
%
%\large
\section{Introduction}
\label{intro}
The hypotheses~\cite{arkani,RS} that there exist extra spatial dimensions with
length scales large enough to be probed by the Large Hadron Collider (LHC)
lead to the possibility that quantum gravity can be investigated via the production
and detection of microscopic black holes~\cite{bhlhc,CH,BHreview}.
In particular, in the Randall-Sundrum (RS) brane-world of Ref.~\cite{RS},
our world is a four-dimensional brane (with coordinates $x^\mu$, $\mu=0,\ldots,3$)
embedded in a five-dimensional manifold whose metric, without other sources present,
is given by
\be
\ud s^2 =e^{-\kappa\,|y|}\,g_{\mu\nu}\,\ud x^{\mu}\,\ud x^{\nu} + \ud y^2
\ ,
\ee
where $y$ parameterizes the fifth dimension.
In the above, $\kappa$ is a deformation parameter, with units of inverse length,
determined by the brane tension (for bounds on $\kappa$, see, e.g.,~\cite{harko}
and References therein).
It also relates the four-dimensional Planck mass $\mpl$ to the five-dimensional
gravitational mass $M_{(5)}$ and one can thus have $M_{(5)}\simeq 1\,$TeV$/c^2$.
This, in turn, allows for the existence of black holes with mass in the TeV~range.
In order to be phenomenologically viable, the brane must also have a thickness,
which we denote by $L$, below which deviations from the four-dimensional
Newton law occur.
Current precision experiments require that $L \lesssim 44\, \mu$m~\cite{Lbounds}.
In the analysis below, the parameters $\kappa$ and $L$ are assumed to be
independent of one another.
\par
Describing black holes in the presence of extra dimensions (for some reviews,
see Refs.~\cite{BHreview}) has proven a rather difficult and stimulating topic.
In fact, to date, only approximate black hole metrics are known on the
brane~\cite{dadhich,bwbh}.
Since the Standard Model interactions are confined to the brane,
and gravity is the only force which acts in the bulk as a whole,
when a black hole decays, the decay products are confined
to the brane except for gravitons.
In a previous publication~\cite{CH}, we showed that, using the metric
of Ref.~\cite{dadhich}, and depending upon the choice of parameter
values, black hole lifetimes can be very long.
If the RS~model is a valid representation of the physical world,
then the black holes created on the brane can live long enough to escape the
LHC and penetrate into the Earth.
As they travel through matter, the black holes can accrete by absorbing nuclei,
electrons or any other matter which comes within their capture radii~\cite{giddings}.
A conjecture has been made~\cite{plaga} that, for the longer-lived black holes
predicted by the model discussed in~\cite{CH}, accretion by the black holes
might bring them to rest within the Earth and allow them to grow to sizes at
which their radiation would be catastrophic.  
This conjecture was criticized in Ref.~\cite{GM2}, where it was argued
that the black hole energy release conjectured in~\cite{plaga} was greatly
overestimated.
In this paper we analyze this conjecture by solving the system of equations
which describes the mass of a black hole and its momentum as functions
of time for various initial conditions and various values of the parameters
which occur in the model developed in Ref.~\cite{CH}.
Based on the results of these calculations, we comment on the possibility of
catastrophic black hole growth on Earth within the RS~scenario.
\par
We shall use units with $1=c=\hbar=\mpl\,\lp$,
where $\mpl\simeq 2.2\cdot 10^{-8}\,$kg and $\lp\simeq 1.6\cdot 10^{-35}\,$m
are the Planck mass and length related to the
four-dimensional Newton constant $G_{\rm N}=\lp/\mpl$.
The corresponding constants in $D=4+d$ dimensions are denoted
by $M_{(D)}$ and $\ell_{(D)}$, respectively.
In our analysis we shall consider only the $D=5$ dimensional RS scenario
with $M_{(5)}\simeq \mew\simeq 1\,$TeV ($\simeq 1.8\cdot 10^{-24}\,$kg),
the electro-weak scale, and $\ell_{(5)}\simeq 2.0\cdot 10^{-19}\,$m.
\section{Black Hole Decay}
\label{decay}
In order to determine the black hole mass as a function of time, the accretion and
decay rates must be expressed in terms of the dynamical quantities of the system.
The decay rate is determined by multiplying the luminosity by the horizon area.
The luminosity is calculated by means of either the canonical or microcanonical
ensemble~\cite{CH,bc2,mfd}, depending upon the relative sizes of the black hole
mass $M$ and the energy scale for the emitted particles.
\subsection{Canonical Ensemble}
If the mass of the black hole is much larger than the typical energy scale of the
emitted particles, one can safely introduce a black hole temperature
$T_{\rm H}=\beta_{\rm H}^{-1}$ and the canonical ensemble is then appropriate to calculate
the luminosity.
The luminosity per unit area in $D$ space-time dimensions in this case is given by
\be
{\cal L}_{(D)}
\simeq
\sum_s\,\int_0^\infty \Gamma_s\,
\frac{\omega^{D-1}\,\ud\omega}
{e^{\beta_{\rm H}\,\omega}\mp 1}
=
f_{(D)}\,T_{\rm H}^{D}
\ ,
\label{L}
\ee
where $\Gamma_s$ denotes grey-body factors and $f$ is a coefficient which depends
upon the number of available particle species $s$ with energy smaller than the
Hawking temperature $T_{\rm H}$.
In general, the decay rate of the black hole is then given by
\be
\frac{\ud M}{\ud\tau}=-{\cal A}_{(D)}\, {\cal L}_{(D)}
\ ,
\ee
where $\tau$ is the black hole proper time, ${\cal A}_{(D)}=S_{(D)}\,\rh^{D-2}$
the horizon area, $S_{(D)}$ the area of a unit sphere, and $\rh$ the
horizon radius in $D$ space-time dimensions.
For example, for a $D=4$ Schwarzschild black hole the relevant energy scale is
the Planck mass $\mpl$ and the well known result~\cite{hawking}
\be
\frac{\ud M}{\ud\tau}\simeq f_{(4)}\,\frac{\mpl}{\lp}\,\left(\frac{\mpl}{M}\right)^2
\ ,
\label{L4>}
\ee
is obtained when the black hole mass $M \gg \mpl$.
\subsection{Microcanonical Ensemble}
When the black hole mass $M$ is on the order of the energy scale for the
emitted particles, the appropriate ensemble to use is the microcanonical
ensemble.
The occupation number density for the Hawking particles in the microcanonical
ensemble is given by~\cite{bc2}
\be
n_{(D)}(\omega)=B\sum_{l=1}^{[[M/\omega]]}\,
\frac{\exp\left\{4\,\pi\,\left(\frac{M-l\,\omega}{M_{(D)}}
\right)^{\frac{D-2}{D-3}}\right\}}
{\exp\left\{4\,\pi\,\left(
\frac{M}{M_{(D)}}\right)^{\frac{D-2}{D-3}}
\right\}}
\ ,
\label{n}
\ee
where $[[X]]$ denotes the integer part of $X$ and $B=B(\omega)$
encodes deviations from the area law~\cite{r1} (in the following we
shall also assume $B$ is constant in the range of interesting values of
$M$).
In the limit $M \to \infty$, $n_{(D)}(\omega)$ mimics the canonical
ensemble (Planckian) number density and Eq.~\eqref{L} is recovered.
\subsection{Metric for a RS Black Hole}
Since gravity propagates in the bulk, a black hole confined to a brane will
produce a back-reaction on the brane itself.
This effect can likely be described in the form of a tidal ``charge'' $q$, and
the effective four-dimensional metric for such a system is thus given
by~\cite{dadhich}
\be
\ud s^2 =
- A\,\ud t^2 + A^{-1}\,\ud r^2 + r^2\left(\ud\theta^2 +\sin(\theta)^2\,\ud\phi^2\right)
\ ,
\label{tidal}
\ee
where
\be
A=1-\frac{2\,M\, \lp}{\mpl\,r}-\frac{q\,\mpl^2\,\lp^2}{M_{(5)}^2\,r^2}
\ ,
\ee
with $M_{(5)}\simeq \mew$ the fundamental mass scale.
A dimensional analysis (and plausibility arguments) shows that the
(dimensionless) tidal charge $q$ must depend upon the black hole
mass and, at least for $M\sim \mew$, it is given by
\be
q \sim \left(\frac{\mpl}{\mew}\right)^\alpha\frac{M}{\mew}
\ .
\ee
In the following analysis we shall assume that $\alpha = 0$~\footnote{This
choice was first made in Ref.~\cite{CH}.
Other cases will be considered elsewhere.}.
Since $\mew \simeq 1\,$TeV$/c^2$, the tidal term in the metric dominates
over the usual General Relativistic $1/r$ term for black hole masses
up to extremely high values.
The range of values for which the tidal term dominates is determined by
the critical mass parameter, $\mc$, which is the mass analogue of
the length $L$  mentioned in the Introduction and which we discuss below.
In this range of values the outer horizon radius is given by
\be
\rh \simeq \lp\,\frac{\mpl}{M_{(5)}}\,\sqrt{\frac{M}{M_{(5)}}}
\ ,
\label{R5}
\ee
which, for $M>M_{(5)}\simeq \mew$, is larger than the
usual four-dimensional expression
\be
R_{\rm h} = 2\,\lp\,\frac{M}{\mpl}
\ .
\label{RH}
\ee
\par
The number density used to calculate the luminosity is determined by the
effective four-dimensional Euclidean action~\cite{CH,gergely},
\be
S_{(4)}^{\rm E}
=
\frac{\mpl}{16\,\pi\,\lp}\,4\,\pi\,\rh^2
=
\frac{\mpl\,\lp\,M}{4\,\meff}
\ ,
\label{SE}
\ee
where the effective energy scale $\meff$ is defined by
\be
\meff = \left(\frac{\mew}{\mpl}\right)^2\mew
\ .
\ee
Since $\mew\ll\mpl$, the effective energy scale is very small compared to
the electro-weak energy scale near which the black holes are created.
For black hole masses much larger than the effective energy scale
the microcanonical and canonical ensembles give the same expression
for the luminosity.
\par
Using Eq.~\eqref{SE}, we showed in Ref.~\cite{CH} that the decay rate can
be written as
\be
\frac{\ud M}{\ud\tau}=
-C\left(\frac{\mpl}{\mew}\right)^2\frac{M}{\mew}
\ .
\label{evap5}
\ee
where $C$ is a numerical constant which can be obtained by equating the
above decay rate to the four-dimensional Hawking decay rate~\eqref{L4>}
for $M \simeq \mc$, that is
\be
C=\frac{g_{\rm eff}\,\mpl}{960\,\pi\,\lp}
\left(\frac{\mew}{\mc}\right)^3
\ ,
\label{C}
\ee
where $\mc$ is again the critical mass and $g_{\rm eff}$ the number of
degrees of freedom into which the black hole can decay.
\par
The value of $\mc$ depends upon the defining condition used.
One possible choice is to require that the black hole horizon radius
equal the four-dimensional expression~\eqref{RH} for sufficiently large
mass, that is
\be
R_{\rm h}
\simeq L
\ ,
\ee
for $M\simeq \mc$.
In this case, one has
\be
\mc
\simeq
\mpl\left(\frac{L}{\lp}\right)
\equiv
M_{\rm c}^{\rm h}
\ .
\ee
Another possibility is to require that the effective horizon radius~\eqref{R5}
not exceed $L$,
\be
\rh \simeq L
\ ,
\label{bH}
\ee
for $M\simeq \mc$, which yields
\be
\mc
\simeq
\mew\left(\frac{L\, \mew}{\lp\, \mpl}\right)^2
\equiv M_{\rm c}^{\rm H}
\ .
\ee
Since $\mew\ll\mpl$ and $L\lesssim 44\,\mu$m, this represents a stronger
constraint on the possible value of the critical mass, namely
$M_{\rm c}^{\rm H}\ll M_{\rm c}^{\rm h}$.
For example, setting $L\simeq 1\,\mu$m and $\mew=1\,$TeV$/c^2$
gives
\be
M_{\rm c}^{\rm H}
\simeq
10^{26}\,{\rm TeV}/c^2
\simeq 10^2\, {\rm kg}
\ .
\ee
\par
It is important to remark that to a given value of $\mc$ there correspond two different
critical values of the horizon radius, namely $\rh(\mc)\sim\sqrt{\mc}$ and
$R_{\rm h}(\mc)\sim\mc$.
This means that the horizon radius of an accreting~\footnote{The opposite picture
would arise for an evaporating black hole which is four-dimensional
(with initial mass $M>\mc$) and reaches the transition stage from above.}
five-dimensional black hole,
necessarily starting with a mass $M<\mc$, can first be approximated by $\rh$ in
Eq.~\eqref{R5} with $M=M(t)$. 
If $M$ reaches the critical mass $\mc$, the black hole becomes four-dimensional and
the expression of its horizon radius afterwards changes from $\rh$ to $R_{\rm h}$
given in Eq.~\eqref{RH}.
The difference $R_{\rm h}(\mc)-\rh(\mc)$ is always negative (meaning
the horizon should shrink at the transition) and can be rather large in magnitude.
For example, on using the value of $M_{\rm c}^{\rm H}\simeq 10^2\,$kg,
$\rh(M_{\rm c}^{\rm H})\simeq 10^{-6}\,$m
$\gg R_{\rm h}(M_{\rm c}^{\rm H})=10^{-25}\,$m.
Such a huge ($19$ orders of magnitude) jump in the horizon radius likely signals
that we need a better description of RS black holes near the dimensional
transition scale (that is, for $M\simeq\mc$).
However, we shall show in Section~\ref{evo} that $M\ll\mc$ at all times for black
holes produced at the LHC, and the approximation outlined above is therefore
adequate for the present analysis.
\par
In any case, the radius of the five-dimensional black hole cannot exceed the
current experimental limit on the size of corrections to Newton law's, that
is $\rh(\mc)\ll 44\,\mu$m, which, for $M_{(5)}\simeq \mew=1\,$TeV$/c^2$,
implies $\mc\ll 10^3\,$kg (from Eq.~\eqref{R5})~\footnote{Larger values
of $\mc$ would be allowed for $M_{(5)}\gg \mew$, but then black holes could
not be produced at the LHC.}.
In the following, we shall provide an argument which actually places
a stronger bound on $\mc$, namely $\mc\ll M_{\rm c}^{\rm H}\simeq 10^2\,$kg.
\section{Black Hole Accretion}
\label{accretion}
After the black holes are created at the LHC they can, depending on the value
of $\mc$,
live long enough in the RS scenario to escape into the atmosphere or into the Earth.
They can grow in mass and therefore in horizon radius by absorbing anything which
comes within their capture radii.
There are two basic mechanisms by which the black holes in general might accrete:
one due to their collisions with the atomic and sub-atomic particles they encounter
as they sweep through matter,
and one due to the gravitational force the black holes exert on
surrounding matter once they come to rest.
The latter form is known as Bondi accretion and is appreciable only when the
black holes have horizon radii greater than atomic size.
\subsection{Capture Radius}
\label{IIIA}
The accretion rate due to collisions is given by
\be
\left.\frac{\ud M}{\ud t}\right|_{\rm acc}
=
\pi\,v\,\rho\,\reff^2
\ ,
\label{acc}
\ee
where $\rho$ is the density of the material through which the black hole is moving,
and $v$ is the relative velocity of the black hole and the surrounding matter,
while $t$ is the time of observers at rest with respect to the medium.
The effective radius $\reff$ depends upon the details of the accretion mechanism
and, for the problem at hand, of the extra-dimensional scenario.
\par
For sufficiently small horizon radius $\rh$, the capture radius $\reff$ can
be determined by simple Newtonian arguments.
In particular, we can assume that it is given by the range over which the
gravitational force of the black hole can overcome
the electromagnetic force which binds the nucleus of an atom to the surrounding
medium.
For a black hole in motion through matter, accretion will then occur when the impact
parameter is small enough for the gravitational field of the black hole to overcome
the electromagnetic binding force.
\par
An expression for the electromagnetic capture radius, $R_{\rm EM}$, in $D$ dimensions
was obtained in Ref.~\cite{giddings} and reads
\be
\rem
=\lp\,\frac{\mpl}{M_{(D)}}
\left(\frac{\beta_D\, M}{M_{(D)}}\right)^{1/(D-1)}
\ ,
\label{REM}
\ee
where
\be
\beta_D = \frac{(D-1)^{D-1}}{(D-2)^{D-2}}\frac{\bar{k}_D\, M_{(D)}^2\, m}{\mpl^2\,\lp^2\,K}
\ ,
\ee
$\bar{k}_D$ and $K$ are constants and $m$ is the mass of the absorbed nucleus.
For $D=5$ and using Eq.~\eqref{R5}, we can rewrite the capture radius as
\be
\rem=C_{\rm EM}\,M^{1/4}
\ ,
\label{Cem}
\ee
in which $C_{\rm EM}$ is a constant which depends on $K$ and $m$.
\par
Of course the above expressions are meaningful only if $\rem\gg \rh$, otherwise 
the Newtonian argument leading to Eq.~\eqref{REM} fails.
As we shall show in Section~\ref{IVC}, black holes created at the LHC indeed have
$\rh\ll \rem$. 
It will therefore be consistent to set $\reff=\rem$ in the accretion rate~\eqref{acc}
in Section~\ref{evo}.
\par
The above expression for $\rem$ also allows us to restrict the possible values
of $\mc$ even further.
Since Eq.~\eqref{REM} follows from a metric of the form~\eqref{tidal}, it will 
hold only at distances shorter than $L\simeq 44\,\mu$m, that is for
\be
\rem\ll L
\ ,
\ee
which yields the bound
\be
\mc\simeq \left(\frac{L}{C_{\rm EM}}\right)^4
\ .
\label{bEM}
\ee
This is a stronger bound than that following from Eq.~\eqref{bH}
for $L\lesssim C_{\rm EM}^2\,\mew/\lp\,\mpl$.
For example, with $L\simeq 1\,\mu$m, $D = 5$,
$M_{(5)}\simeq\mew$, $\bar k=2/3\,\pi$, $K=224\,$J$/$m$^2$
and $m\simeq 9\cdot 10^{-27}\,$kg, one obtains
\be
C_{\rm EM}\simeq 1.1\cdot 10^{-6}\,{\rm m}/{\rm kg}^{1/4}
\ ,
\label{cem}
\ee
and
\be
\mc\simeq 0.6\,{\rm kg}\ll M_{\rm c}^{\rm H}
\ .
\ee
\par
In order to properly describe accretion, we next need to consider two relevant
length scales, namely the typical size of an atom, that is $1\,$\AA, and the length
$L\lesssim 44\,\mu$m associated with extra-dimensional corrections
to Newton's law.
Since we assume $L\gtrsim 1\,$\AA, a black hole of subatomic size will always be
described as five-dimensional.
\subsection{Subatomic Accretion}
For $\rh\ll\rem\ll 1\,$\AA, the relevant accretion rate is given by Eq.~\eqref{acc}
with $\reff\simeq\rem$.
When such a subatomic black hole absorbs a nucleus or electrons,
it acquires the charge and spin of the absorbed particles as well as their masses.
In this analysis the assumption is made that the spins of the absorbed particles are
randomly directed so there is no net angular momentum of the black hole as
it accretes.
Also the charge on the absorbed nuclei is assumed to be neutralized by the absorption
of valence electrons, which will be drawn into the black hole by the Coulomb force until
the black hole is neutralized.
A further assumption is that the Earth's density is uniform, although the conclusions
would not be significantly altered by considering a more accurate model of the
Earth's interior.
\par
For the capture radius $\rem\simeq 1\,$\AA, the mass of the five-dimensional
black hole is on the order of kg's [from Eqs.~\eqref{Cem} and~\eqref{cem}].
If the black hole were to reach this size before traversing the Earth's diameter,
most likely, it would have ceased moving and begun accreting by absorbing
any matter which came within its horizon radius.
Further, in the RS~scenario, a black hole with such a mass has an horizon
radius of $\rh \simeq 10^{-6}\,$m [from Eq.~\eqref{R5}], which
is on the order of the assumed thickness of the brane, $L$.
The black hole will therefore accrete like a four-dimensional one beyond
this point.
\subsection{Bondi Accretion}
If a black hole comes to rest inside the Earth, it will continue to accrete according to
the same basic formula as in Eq.~(\ref{acc}).
Now, however, the relative velocity is due to the motion of particles in the surrounding
medium.
A particle whose velocity $v$ is less than the escape velocity at a particular distance
$\reff$ from the black hole will be absorbed.
The accretion rate can thus be written as
\be
\left.\frac{\ud M}{\ud t}\right|_{\rm acc}
=\frac{4\, \pi\,\rho\,\lp^2}{v^3}\left(\frac{M}{\mpl}\right)^2
\ .
\label{accr}
\ee
This type of accretion is known as Bondi accretion (see~\cite{BH, Bondi} and References
therein) and holds for any massive object, e.g.~a star,  which is accreting any surrounding
material which is free to move. 
As will be shown below, Bondi accretion never becomes effective for all values of
the critical mass parameter and initial conditions of interest for the LHC.
\section{Black Hole Evolution}
\label{evo}
The growth and decay of microscopic black holes created at the LHC are described
by the expressions given above for the evaporation rate and the accretion rate along
with the expression for the rate of change of the black hole's momentum.
The solution of this system of equations gives the time evolution of the mass and
momentum~ \footnote{The actual decay
of a black hole is a discrete process which causes jumps in both $M$ and
$p$~\cite{bhlhc}.
The evolution equations~\eqref{dMdt} and~\eqref{dpdt}, as well as the continuous
solutions displayed in the figures, represent valid approximations only for $M$ sufficiently
larger than the mass of the decay products.
The actual decay time is also affected by these considerations.}.
\subsection{System of Equations}
The time evolution of the black hole mass is obtained by summing the evaporation
and accretion expressions,
\be
\frac{\ud M}{\ud t} =
\left.\frac{\ud M}{\ud t}\right|_{\rm evap}
+\left. \frac{\ud M}{\ud t}\right|_{\rm acc}
\ .
\label{dMdt}
\ee
The decay rate in the reference frame of the Earth is obtained from
Eq.~\eqref{evap5},
\be
\left.\frac{\ud M}{\ud t}\right|_{\rm evap}
\simeq
-\frac{1}{\gamma}
\left.\frac{\ud M}{\ud \tau}\right|_{\rm evap}
\ ,
\label{evap}
\ee
where $\gamma$ is the relativistic factor for a point-particle of mass $M$
and three-momentum of magnitude $p$,
\be
\gamma = \frac{1}{\sqrt{1-\beta^2}} = \frac{\sqrt{M^2 + p^2}}{M}
\ .
\ee
Inserting this expression for $\gamma$ and that for $C$ in terms of $\mc$
[see Eq.~\eqref{C}] into Eq.~\eqref{evap} gives
\be
\left.\frac{\ud M}{\ud t}\right|_{\rm evap}
\simeq
-\frac{g_{\rm eff}}{960\, \pi\,\lp}\left(\frac{\mpl}{\mc}\right)^3
\frac{M^2(t)}{\sqrt{M^2(t)+p^2(t)}}
\ .
\ee
The accretion rate is given in general by Eq~\eqref{acc}.
For sub-atomic growth and $p>0$, the accretion rate is
\be
\left.\frac{\ud M}{\ud t}\right|_{\rm acc} = \pi\, \rho\, v(t)\, \rem^2
\ ,
\ee
where $\rho\simeq 5.5\cdot 10^3\,$kg$/$m$^3$ is the Earth's mean density and
$\rem$ is given by Eq.~\eqref{Cem} expressed in terms of a time-dependent
mass $M=M(t)$.
\par
Finally, the time-evolution of the momentum in the Earth frame is described by the equation
\be
\frac{\ud p}{\ud t}
&\!\!=\!\!&
\frac{p(t)}{M(t)}\left.\frac{\ud M}{\ud t}\right|_{\rm evap}
\nonumber
\\
&\!\!\simeq\!\!&
-\frac{g_{\rm eff}}{960\, \pi\,\lp} \left(\frac{\mpl}{M_c}\right)^3
\frac{M(t)\,p(t)}{\sqrt{M^2(t)+p^2(t)}}
\ .
\label{dpdt}
\ee
\par
The net change of mass with respect to time~\eqref{dMdt} and the
equation~\eqref{dpdt} for the time evolution of the momentum form
a system of equations which can be solved numerically to obtain $M(t)$ and $p(t)$.
Note that accretion dominates only if the momentum is larger than the critical value
\be
p_{\rm c}
=\frac{g_{\rm eff}\,\mew^{3/2}}{960\,\pi^2\,\lp\,\rho\,C_{\rm EM}^2}
\left(\frac{\mpl}{\mc}\right)^3
\left(\frac{M}{\mew}\right)^{3/2}
,
\label{pcr}
\ee
in which we again used $\reff=\rem$.
\subsection{Time evolution of mass and momentum}
\label{IVB}
\begin{figure}[t!]
\centering
\raisebox{3cm}{$M$}
\epsfxsize=2.9in
\epsfbox{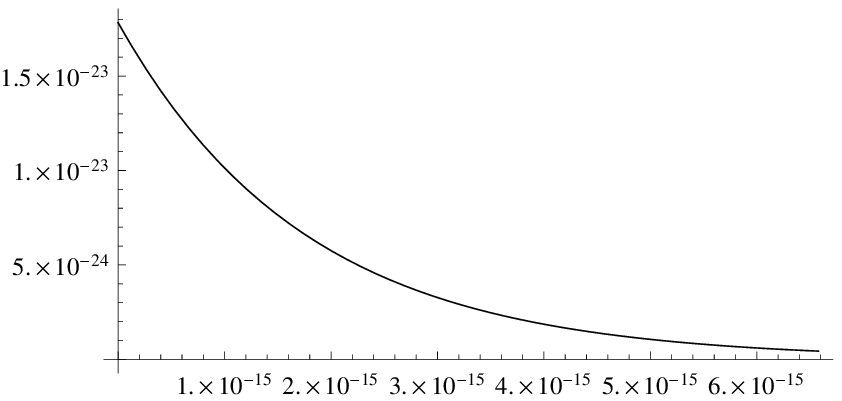}
\\
{\hspace{2in} $t$}
\\
\qquad
\raisebox{3cm}{$p$}
\epsfxsize=2.7in
\epsfbox{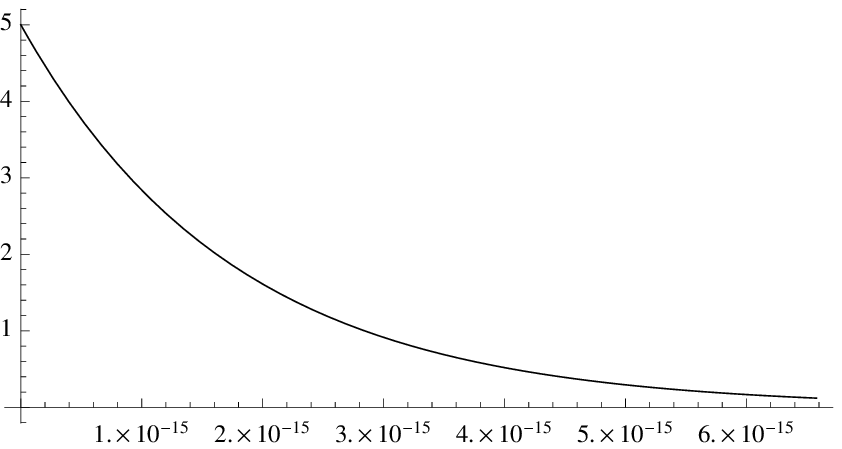}
\\
{\hspace{2in} $t$}
\caption{Mass (in kg; upper panel) and momentum in (TeV$/c$; lower panel)
of a black hole as a function of time (in sec) when $p(0)<p_{\rm c}(0)$.}
\label{Mt2}
\end{figure}
\begin{figure}[t!]
\centering
\raisebox{3cm}{$M$}
\epsfxsize=2.9in
\epsfbox{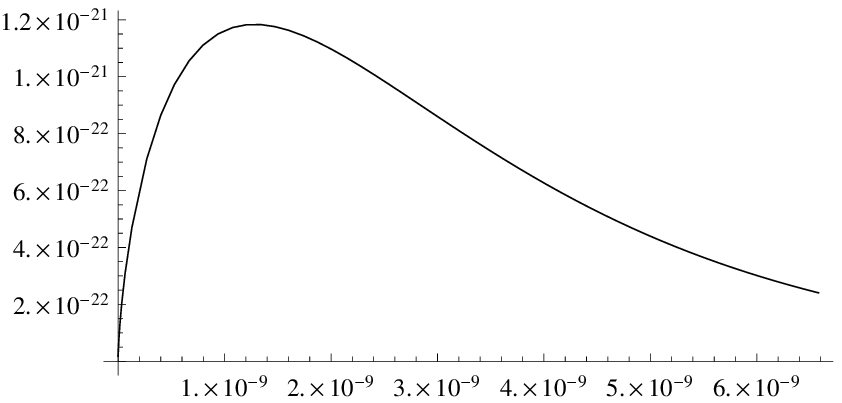}
\\
{\hspace{2in} $t$}
\\
\qquad
\raisebox{3cm}{$p$}
\epsfxsize=2.7in
\epsfbox{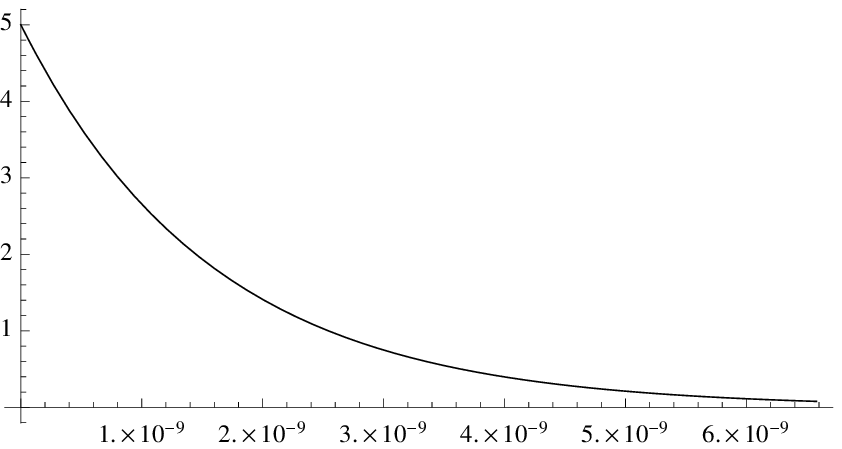}
\\
{\hspace{2in} $t$}
\caption{Mass (in kg; upper panel) and momentum (in TeV$/c$; lower panel)
of a black hole as a function of time (in sec) when $p(0)>p_{\rm c}(0)$.}
\label{Mt6}
\end{figure}
The solutions to the above system of equations show a rather simple qualitative
behavior:
for a given value of the critical mass $\mc$, there are initial conditions for which the
black hole never accretes ($p<p_{\rm c}$ from the outset).
Otherwise, it first accretes and then begins to evaporate, since, for large black hole mass,
the evaporation rate in the laboratory frame grows like $M$ whereas the accretion rate
decreases like $M^{-1/2}$.
\par
A typical example of the first kind of evolution for the mass and momentum
is shown in Fig.~\ref{Mt2} for a black hole with initial mass
$M(0)=10\,$TeV$/c^2\simeq 1.8\cdot 10^{-23}\,$kg and momentum
$p(0) = 5\,$TeV$/c\simeq 2.7\cdot 10^{-15}\,$kg$\,$m$/$sec~\footnote{These
values correspond to a black hole energy of
about $11\,$TeV in the laboratory  and were chosen considering the LHC total collision energy
of $14\,$TeV and the fact that a black hole cannot be the only product of a collision.},
with $\bar k=2/3\,\pi$, $K=224\,$J$/$m$^2$ and $m=9.0\cdot 10^{-26}\,$kg
(yielding $C_{\rm EM}\simeq 1.1\cdot 10^{-6}\,$m$/$kg$^{1/4}$,
$\rh(0)\simeq 6.1\cdot 10^{-19}\,$m and $\rem(0)\simeq 2.3\cdot 10^{-12}\,$m)
and $\mc = 10\,$kg.
With this choice of parameters, $p(0)<p_{\rm c}(0)$ and the black hole
just evaporates.
Note, though, that the mass plot does not resemble the usual Hawking
behavior but rather a more conventional (quasi-exponential) decay,
with a much longer decay time~\cite{CH}.
\par
The second kind of evolution ($p(0)>p_{\rm c}(0)$)
is displayed in Fig.~\ref{Mt6}, with the same initial
mass and momentum and $\mc= 10^3\,$kg.
The maximum mass $\mmax\simeq 1\cdot 10^{-21}\,$kg
is reached about $1\cdot 10^{-9}\,$sec after production and
corresponds to a horizon radius $\rh\simeq 5\cdot 10^{-18}\,$m
and capture radius $\rem\simeq 7\cdot 10^{-12}\,$m.
Subsequently, evaporation dominates and the decay-time of such a black hole
is about $3\cdot 10^{-7}\,$sec, corresponding to
a travelled distance of around $7\cdot 10^{-3}\,$m.
Note that the momentum behaves the same in both cases, but the time scales are
significantly different.
\subsection{Impact of the Critical Mass Scale}
\label{IVC}
\begin{figure*}[t]
\centering
\raisebox{3cm}{$\log_{10}(\mmax)$}
\epsfxsize=2.6in
\epsfbox{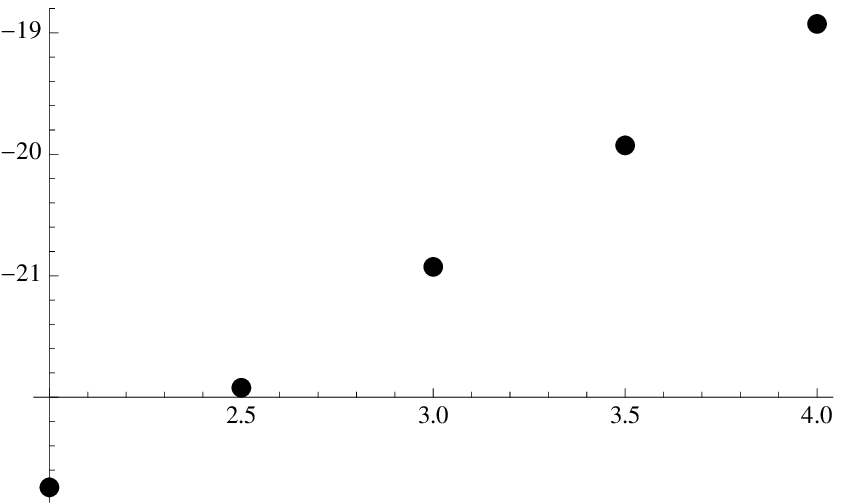}
\qquad
\raisebox{3cm}{$\log_{10}(\rem)$}
\epsfxsize=2.6in
\epsfbox{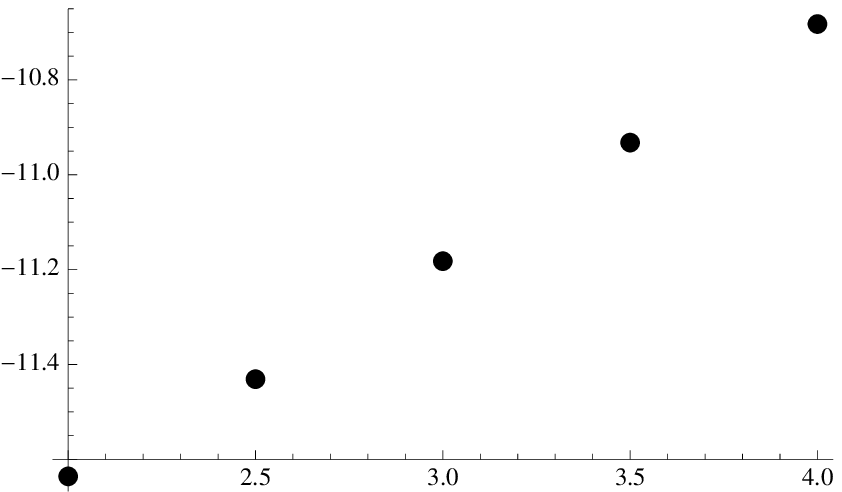}
\qquad
\\
$\log_{10}(\mc)$
{\hspace{6cm} $\log_{10}(\mc)$}
\caption{Black hole maximum mass (in kg; left panel) and maximum 
capture radius (in m; right panel) as a function of the critical mass $\mc$
for growing black holes ($10^2\,$kg $\le\mc\le 10^4\,$kg, $M(0)=10\,$TeV$/c^2$
and $p(0)=5\,$TeV$/c$).}
\label{sM}
\end{figure*}
\begin{figure*}[t]
\centering
\raisebox{3cm}{$\log_{10}(t)$}
\epsfxsize=2.6in
\epsfbox{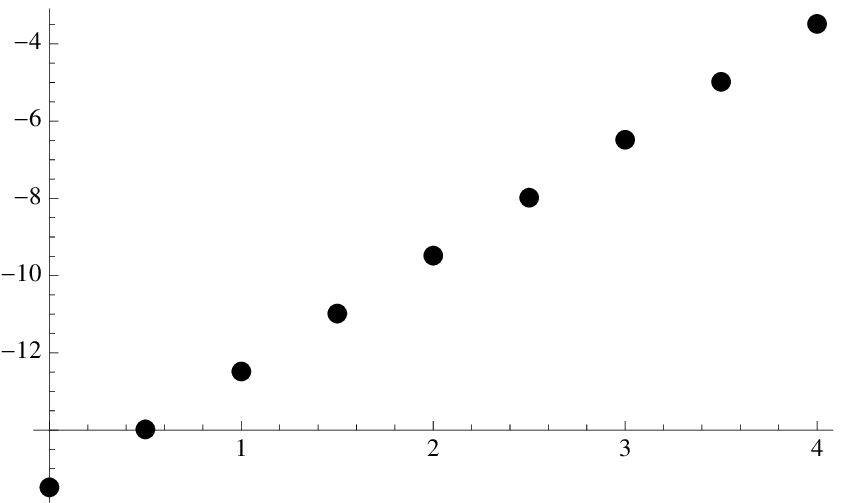}
\qquad
\raisebox{3cm}{$\log_{10}(T)$}
\epsfxsize=2.6in
\epsfbox{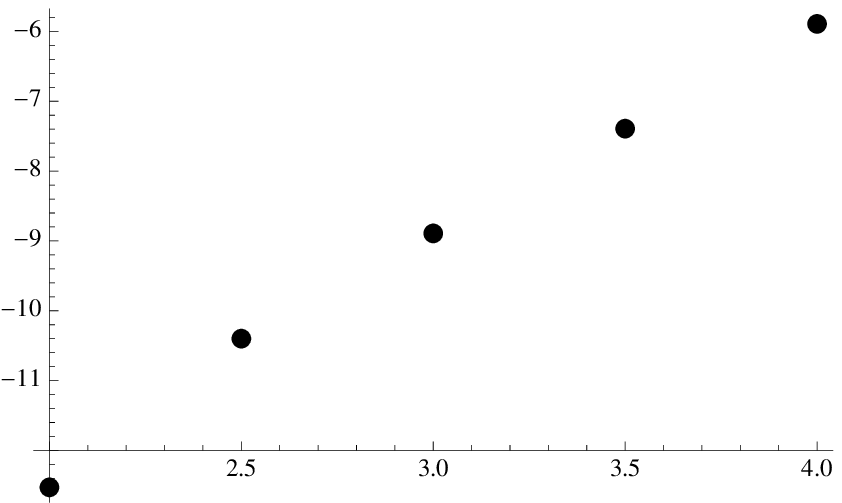}
\qquad
\\
$\log_{10}(\mc)$
{\hspace{6cm} $\log_{10}(\mc)$}
\caption{Decay time (in sec; left panel, for $1\,$kg $\le\mc\le 10^4\,$kg)
and time to mass peak
(in sec; right panel, for $10^2\,$kg $\le\mc\le 10^4\,$kg) as a function of
the critical mass $\mc$ ($M(0)=10\,$TeV$/c^2$ and $p(0)=5\,$TeV$/c$).}
\label{sT}
\end{figure*}
\begin{figure*}[bh!]
\centering
\raisebox{3cm}{$\log_{10}(s)$}
\epsfxsize=2.6in
\epsfbox{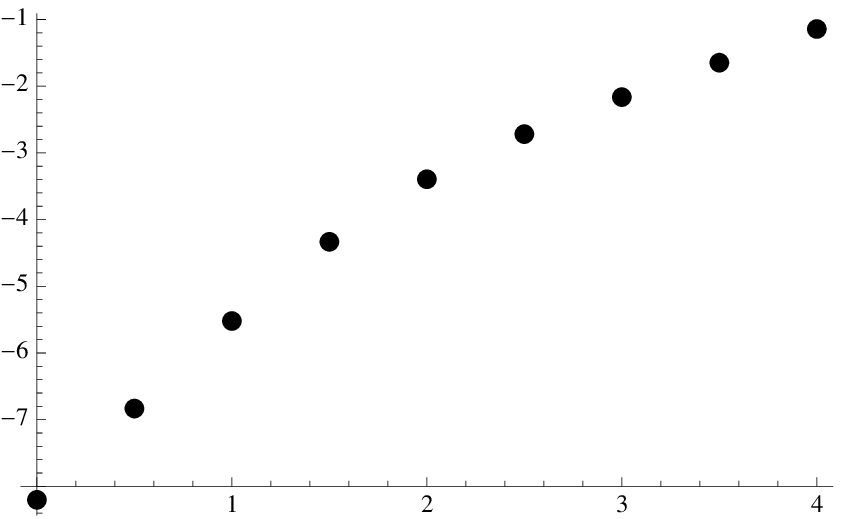}
\qquad
\raisebox{3cm}{$\log_{10}(S)$}
\epsfxsize=2.6in
\epsfbox{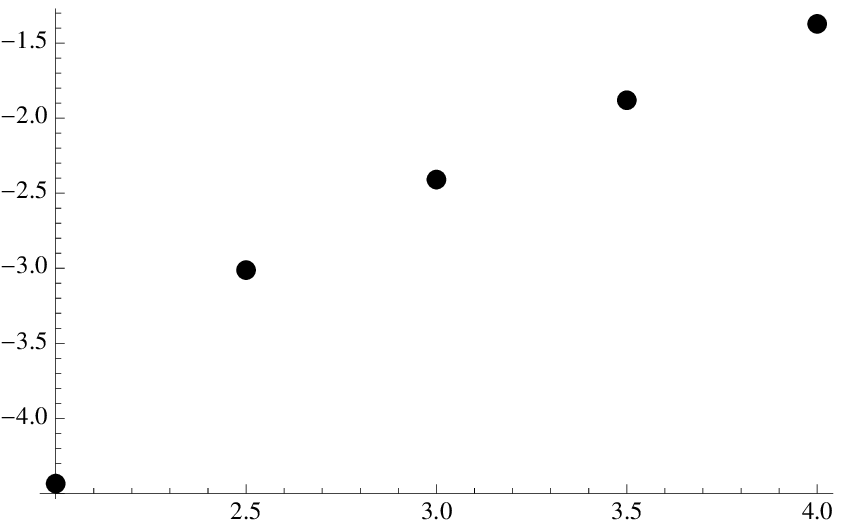}
\qquad
\\
$\log_{10}(\mc)$
{\hspace{6cm} $\log_{10}(\mc)$}
\caption{Maximum travelled distance (in m; left panel, for $1\,$kg $\le\mc\le 10^4\,$kg)
and travelled distance to the mass peak
(in m; right panel, for $10^2\,$kg $\le\mc\le 10^4\,$kg) as a function of
the critical mass $\mc$ ($M(0)=10\,$TeV$/c^2$ and $p(0)=5\,$TeV$/c$).}
\label{sS}
\end{figure*}
Since the choice of the critical mass is not unique, it is of interest to study the effect
of the value of the critical mass $\mc$ on the essential quantities associated with the
time evolution of a black hole produced at the LHC.
\par
For the same initial conditions $M(0)=10\,$TeV$/c^2$ and $p(0)= 5\,$TeV$/c$
used in Fig.~\ref{Mt6},
one can evolve the system considering different values of $\mc$ (see Figs.~\ref{sM}
and~\ref{sS}).
Although we showed in previous Sections that a reasonable upper bound for
$\mc\simeq 1\,$kg, we evolved the system in the range $1\,$kg $\le\mc\le 10^4\,$kg.
One first finds that the black hole  accretes ($p(0)>p_{\rm c}(0)$) only when
$\mc\gtrsim 10^2\,$kg.
The maximum black hole mass then very closely follows the scaling law
(see Fig.~\ref{sM})
\be
\mmax
\propto
\mc^2
\ ,
\ee
for $\mc> 10^2\,$kg.
Note that $\mmax\ll \mc$ and the black holes remain five-dimensional
in all cases considered.
According to Eq.~\eqref{Cem}, the above scaling law implies that a
similar law also holds for the maximum value of the capture radius, that is
\be
\rem
\propto
\mc^{1/2}
\ .
\ee
The decay time $t$ and and the time $T$ to reach $\mmax$ are shown in Fig.~\ref{sT},
the total travelled distance $s$ and the distance $S$ travelled
to reach $\mmax$ are shown in Fig.~\ref{sS}.
Note that $s$ grows more slowly with $\mc$ for the cases in which accretion
is significant.
\subsection{Impact of the Initial Conditions}
\begin{table}[th!]
\centering
\begin{tabular}{|c| c| c| c| c| c|}
\hline
$\mc$  (kg) & $\mmax$  (kg)& $\rem$ (m) & $\rh$ (m) & $S$ (m) & $T$ (sec)
\\
\hline
$10^2$
& $1.2\cdot 10^{-23}$ & $2.1\cdot 10^{-12}$ & $5.1\cdot 10^{-19}$ & $3.1\cdot 10^{-4}$ & $1.7\cdot 10^{-12}$ 
\\
$10^3$
& $1.2\cdot 10^{-21}$ & $6.6\cdot 10^{-12}$ & $5.1\cdot 10^{-18}$ & $4.2\cdot 10^{-3}$ & $1.3\cdot 10^{-9}$  
\\
$10^4$
& $1.2\cdot 10^{-19}$ & $2.1\cdot 10^{-11}$ & $5.1\cdot 10^{-17}$ & $4.3\cdot 10^{-2}$ & $1.3\cdot 10^{-6}$ 
\\
\hline
\end{tabular}
\caption{Data in this table are for initial conditions:
$M(0)=1\,$TeV$/c^2$ ($=1.8\cdot 10^{-24}\,$kg)
and $p(0)=5\,$TeV$/c$.
\label{T1}}
\end{table}
\begin{table}[ht!]
\centering
\begin{tabular}{| c| c| c| c| c| c|}
\hline
$\mc$ (kg)& $\mmax$  (kg)& $\rem$ (m) & $\rh$ (m) & $S$ (m)& $T$ (sec)
\\
\hline
$10^2$
& $1.8\cdot 10^{-23}$ & $2.3\cdot 10^{-12}$ & $6.3\cdot 10^{-19}$ & $3.7\cdot 10^{-5}$ & $3.0\cdot 10^{-13}$
\\
$10^3$
& $1.2\cdot 10^{-21}$ & $6.5\cdot 10^{-12}$ & $5.0\cdot 10^{-18}$ & $3.9\cdot 10^{-3}$ & $1.3\cdot 10^{-9}$
\\
$10^4$
& $1.2\cdot 10^{-19}$ & $2.1\cdot 10^{-11}$ & $5.1\cdot 10^{-17}$ & $4.3\cdot 10^{-2}$ & $1.3\cdot 10^{-6}$
\\
\hline
\end{tabular}
\caption{Data in this table are for the initial conditions:
$M(0)=10\,$TeV$/c^2$ ($=1.8\cdot 10^{-23}\,$kg) and $p(0)=5\,$TeV$/c$.
\label{T2}}
\end{table}
\begin{table}[ht!]
\centering
\begin{tabular}{| c| c| c| c| c| c|}
\hline
$\mc$ (kg)& $\mmax$  (kg)& $\rem$ (m) & $\rh$ (m) & $S$ (m)& $T$ (sec)
\\
\hline
$3\cdot 10^2$
& $4.3\cdot 10^{-23}$ & $2.9\cdot 10^{-12}$ & $9.7\cdot 10^{-19}$ & $4.1\cdot 10^{-4}$ & $3.6\cdot 10^{-11}$
\\
$10^3$
& $4.1\cdot 10^{-22}$ & $5.0\cdot 10^{-12}$ & $3.0\cdot 10^{-18}$ & $2.1\cdot 10^{-3}$ & $1.3\cdot 10^{-9}$
\\
$10^4$
& $4.0\cdot 10^{-20}$ & $1.6\cdot 10^{-11}$ & $3.0\cdot 10^{-17}$ & $2.5\cdot 10^{-2}$ & $1.3\cdot 10^{-6}$
\\
\hline
\end{tabular}
\caption{Data in this table are for the initial conditions:
$M(0)= 10\,$TeV$/c^2$ ($=1.8\cdot 10^{-23}\,$kg) and $p(0)=1\,$TeV$/c$.
\label{T3}}
\end{table}
In order to complete our analysis, we next show the results for different values of
$M(0)$ and $p(0)$. 
In Tables~\ref{T1}--\ref{T3}, $\mmax$ is the maximum value of the mass
attained by the black hole, $\rem$ is the value of the electromagnetic capture
radius and $\rh$ of the horizon radius when $M=\mmax$, $S$ is the distance travelled
by the black hole when it reaches $\mmax$, and $T$ is the time elapsed
before attaining $\mmax$.
The computed quantities are displayed with two digits for convenience,
although it is just the order of magnitude that should be considered.
\par
Comparing the entries between Tables~\ref{T1} and~\ref{T2} shows that,
within the range of values that could be of interest for the LHC,
these quantities do not depend significantly on the initial mass.
On the other hand, lowering the initial momentum increases the value of the
critical mass above which the black holes can grow [since $p_{\rm c}\propto \mc^{-3}$
from Eq.~\eqref{pcr}].
Further, comparing the entries between Table~\ref{T2} and~\ref{T3} shows that
the maximum black hole mass actually decreases for decreasing $p(0)$ at a given
critical mass.
\par
Since the maximum value of the capture radius $\rem$ stays below
$1\,$\AA, even for the physically unreasonable case of $\mc\simeq 10^4\,$kg,
Bondi accretion never becomes effective.
Finally, it is worth noting that, on using the values for the black hole initial velocity
and the maximum values of $\rh$ and $\rem$ given in this Section, one always obtains
$\rh\ll\rem$ for all the allowed initial conditions.
\cleardoublepage
\section{Conclusions}
We have studied the evolution in time of microscopic black holes that could
be produced at the LHC based on our previous paper~\cite{CH} and the
description of brane-world black holes given in Ref.~\cite{dadhich}.
With respect to Ref.~\cite{CH}, accretion has been now included in the
analysis and all of the parameters have been chosen so as to cover a fairly
comprehensive range of possible outcomes.
In particular, our model contains a critical mass scale, $\mc$, which
is related to the transition from the five-dimensional behavior, effectively
described by the metric~\eqref{tidal}, to the usual four-dimensional 
description of a black hole.
\par
As shown in the previous Section, in particular in Tables~\ref{T1}--\ref{T3},
the maximum black hole mass never reaches catastrophic size before
leaving the Earth.
The black hole mass remains at microscopic values for a wide range of
acceptable initial conditions and for a wide range of critical masses, $\mc$.
In order for the black holes created at the LHC to grow at all,
the critical mass should be $\mc\gtrsim 10^2\,$kg.
This value is already larger than the maximum compatible with experimental tests
of Newton's law (and we further relaxed it to $\mc=10^4\,$kg in our analysis).
For smaller values of $\mc$, the black holes cannot accrete fast enough
to overcome the decay rate.
Furthermore, the larger $\mc$ is taken to be, the longer a black hole takes
to reach its maximum value.
\par
The data in Table~\ref{T3} show that, within the warped-brane scenario,
the maximum masses reached by black holes produced at the LHC are about
four orders of magnitude greater than that of a nucleon.
If these black holes or their remnants come to rest in the Earth, they will begin
to Bondi accrete.  
Assuming the extremal conditions that the accreted matter is mostly free iron
nuclei at low energy ($T \approx 300$ K), the rate of Bondi accretion is
[see Eq.~\eqref{accr}] $1.9 \times 10^{-64}\,$kg/s.
For a black hole at rest to accrete even the mass of a single nucleon would
thus require a time interval many orders of  magnitude larger than the age of the
Universe.
\par
We conclude that, for the RS~scenario and black holes described by the
metric~\eqref{tidal}, the growth of black holes to catastrophic size is not possible.
Nonetheless, it remains true that the expected decay times
are significantly longer than is typically predicted by other
models, as was first shown in Ref.~\cite{CH}. 
\section*{Acknowledgments}
R.C.~and B.H.~would like to thank S.~Giddings and R.~Plaga for
many discussions on this topic.
R.C.~would also like to thank R.~Maartens and members of the ICG,
Portsmouth, and O.~Micu and members of the theory group
at Dortmund University. 
\end{document}